\documentclass[11pt]{article}
\usepackage[a4paper]{geometry}
\usepackage{lipsum}
\usepackage[english]{babel}
\usepackage[utf8]{inputenc}
\usepackage{csquotes}
\usepackage{booktabs}
\usepackage{color,graphicx,psfrag,epsf}
\usepackage{enumerate}
\usepackage{amsmath,amsthm,amssymb,amsfonts}
\usepackage{bm,bbm,array}
\usepackage{latexsym}
\usepackage{blkarray}
\usepackage{enumerate}
\usepackage{float}
\usepackage{tikz-qtree,tikz-qtree-compat}
\usepackage{multirow}
\usepackage{setspace}
\usepackage[referable]{threeparttablex}
\usepackage{authblk}
\usepackage[hidelinks]{hyperref}
\usepackage{algorithm}
\usepackage[noend]{algpseudocode}

\usepackage{xcolor}
\usepackage{lipsum}
\makeatletter
\DeclareRobustCommand{\cc}{%
}
\DeclareRobustCommand{\sc}{%
}
\makeatother

\AtBeginEnvironment{tabular}{\onehalfspacing}

\usepackage{natbib}
\definecolor{pigment}{rgb}{0.2, 0.2, 0.6}
\usepackage{hyperref}
\hypersetup{
  colorlinks   = true, 
  urlcolor     = pigment, 
  linkcolor    = pigment, 
  citecolor   = pigment 
}

\title{\vspace{-1cm}
Quantifying Uncertainty in Transdimensional Markov Chain Monte Carlo Using Discrete Markov Models
}

\author[1]{Daniel W. Heck\thanks{
Daniel W. Heck, Statistical Modeling in Psychology, University of Mannheim, Germany, \href{mailto:heck@uni-mannheim.de}{heck@uni-mannheim.de}.
\newline
R code for all simulations is available at the Open Science Framework at \href{https://osf.io/kjrkz}{https://osf.io/kjrkz}, and the R package \texttt{MCMCprecision} is available at \href{https://CRAN.R-project.org/package=MCMCprecision}{https://CRAN.R-project.org/package=MCMCprecision}.}}
\author[2]{Antony M. Overstall}
\author[3]{Quentin F. Gronau}
\author[4]{Eric-Jan Wagenmakers}

\affil[1]{
Statistical Modeling in Psychology, University of Mannheim
(\href{mailto:heck@uni-mannheim.de}{heck@uni-mannheim.de})}
\affil[2]{School of Mathematical Sciences and Southampton Statistical Sciences Research Institute, University of Southampton
(\href{mailto:A.M.Overstall@soton.ac.uk}{A.M.Overstall@soton.ac.uk})}
\affil[3]{Department of Psychology, University of Amsterdam
(\href{mailto:quentingronau@web.de}{quentingronau@web.de})}
\affil[4]{Department of Psychology, University of Amsterdam
(\href{mailto:ej.wagenmakers@gmail.com}{ej.wagenmakers@gmail.com})}

\usepackage{secdot} 
\sectiondot{section}
\sectiondot{subsection}
\sectiondot{subsubsection}
\makeatletter
\renewcommand\section{\@startsection {section}{1}{\z@}%
                                   {-3.5ex \@plus -1ex \@minus -.2ex}%
                                   {2.3ex \@plus.2ex}%
                                   {\centering\normalfont\scshape}}
\renewcommand\subsection{\@startsection {subsection}{1}{\z@}%
                                   {-3.5ex \@plus -1ex \@minus -.2ex}%
                                   {2.3ex \@plus.2ex}%
                                   {\textit}}
\makeatother
\date{}

\begin{document}

\def\spacingset#1{\renewcommand{\baselinestretch}%
{#1}\small\normalsize} \spacingset{1}

\onehalfspacing

\setlength{\skip\footins}{1cm}
\maketitle
\begin{abstract}
Bayesian analysis often concerns an evaluation of models with different dimensionality as is necessary in, for example, model selection or mixture models.
To facilitate this evaluation, transdimensional Markov chain Monte Carlo (MCMC) 
relies on sampling a discrete \cc indexing \sc variable to estimate the posterior model probabilities.
However, little attention has been paid to the precision of these estimates.
If only few switches occur between the models in the transdimensional MCMC output, precision may be low and assessment based on the assumption of independent samples misleading.
Here, we propose a new method to estimate the precision based on the observed transition matrix of the \cc model-indexing \sc variable.
Assuming a first order Markov model, the method samples from the posterior of the stationary distribution.
This allows assessment of the uncertainty in the estimated posterior model probabilities, model ranks, and Bayes factors.
Moreover, the method provides an estimate for the effective sample size of the MCMC output.
In two model-selection examples, we show that the proposed approach provides a good assessment of the uncertainty associated with the estimated posterior model probabilities.

\vspace{.5cm}

\noindent%
\end{abstract}

\doublespacing

\newpage

\section{Introduction}

Transdimensional Markov chain Monte Carlo (MCMC) methods provide an indispensable tool for the Bayesian analysis of models with varying dimensionality \citep{sisson2005transdimensional}. 
An important application is Bayesian model selection, where the aim is to estimate posterior model probabilities $p( \mathcal M_i \mid \bm y)$ for a set of models $\mathcal M_i$, $i=1,\dots,I$ given the data $\bm y$ \citep{kass1995bayes}. 
In order to ensure that the Markov chain converges to the correct stationary distribution, transdimensional MCMC methods such as reversible jump MCMC \citep{green1995reversible} or the product space approach \citep{carlin1995bayesian} match the dimensionality of parameter spaces across different models (e.g., by adding parameters and link functions). 
Transdimensional MCMC methods have proven to be very useful for the analysis of many statistical models including capture-recapture models \citep{arnold2010capturerecapture}, generalized linear models \citep{forster2012reversible}, factor models \citep{lopes2004bayesian}, and mixture models \citep{fruhwirth-schnatter2001markov}, and are widely used in substantive applications such as selection of phylogenetic trees \citep{opgen-rhein2005inference}, gravitational wave detection in physics \citep{karnesis2014bayesian}, or cognitive models in psychology \citep{lodewyckx2011tutorial, heck2017information}.

Crucially, transdimensional MCMC methods always include a discrete parameter $z$ with values in $1,\dots,I$ indexing the competing models. 
At iteration $t=1,\dots,T$, posterior samples are obtained for the indexing variable $z^{(t)}$ and the model parameters, which are usually continuous and differ in dimensionality \citep[for a review, see][]{sisson2005transdimensional}. 
For instance, a Gibbs sampling scheme can be adopted \citep{barker2013bayesian}, in which the \cc indexing variable $z$ \sc and the continuous model parameters are updated in alternating order. 
Such a sampler switches between models depending on the current values of the continuous parameters, and then updates these parameters in light of the current model $\mathcal M_i$ conditionally on the value of $z^{(t)}=i$ \citep{barker2013bayesian}. 
Given convergence of the MCMC chain, the sequence $z^{(t)}$ follows a discrete stationary distribution with probabilities $\bm \pi=(\pi_1,\dots,\pi_I)^\top$. 
Due to the design of the sampler, these probabilities are identical to the posterior model probabilities of interest, $\pi_i = p(\mathcal M_i \mid \bm y)$ and, given uniform model priors $p(\mathcal M_i)= 1/I$, also proportional to the marginal likelihoods $p(\bm y\mid \mathcal M_i)$.
Hence, transdimensional MCMC samplers can be used to directly estimate these posterior probabilities as the relative frequencies of samples $z^{(t)}$ falling into the $I$ categories, $\hat\pi_i =1/T \sum_t \mathbbm I(z^{(t)}=i)$, where $\mathbbm I$ is the indicator function. 
Due to the ergodicity of the Markov chain, this estimator is ensured to be asymptotically unbiased \citep{green1995reversible, carlin1995bayesian}. 

\begin{figure}
\includegraphics[width=\linewidth]{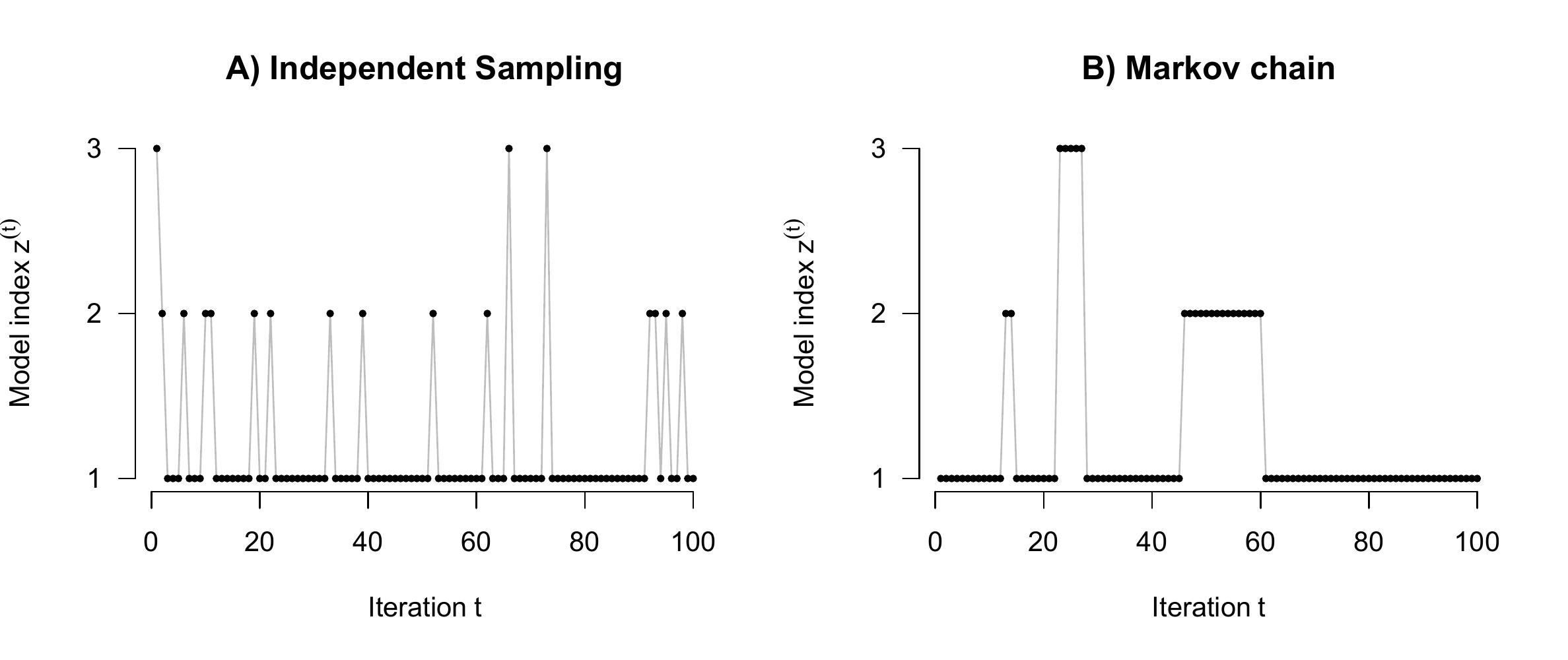}
\caption{
Illustration of $T = 100$ iterations of a discrete model-indexing variable $z^{(t)}$ that were sampled from (A) independent categorical distributions and (B) a Markov model with positive autocorrelation \cc (cf. Section~\ref{s.autocorr}).
Using the method proposed in Section~\ref{s.neff}, the estimated effective sample sizes were $\widehat{T}_\text{eff} = 96$ and $\widehat{T}_\text{eff} = 8$, respectively. \sc
}
\label{f.chain}
\end{figure}

Usually, dependencies due to MCMC sampling are taken into account for continuous parameters \citep{jones2006fixedwidth, flegal2015relative, doss2014markov}. 
In contrast, however, the estimate $\bm {\hat \pi}=(\hat\pi_1,\dots,\hat\pi_I)^\top$  based on the sequence of discrete samples $z^{(t)}$ is usually reported without quantifying estimation uncertainty \cc due to MCMC sampling. \sc
Often, the samples $z^{(t)}$ are correlated to a substantial, but unknown, degree because of infrequent switching between models.
This is illustrated in Figure~\ref{f.chain}, which shows a sequence of independent and correlated samples $z^{(t)}$ in Panels A and B, respectively. 
Inference about the stationary distribution $\bm\pi$ is more reliable in the first case compared to the second case, in which the autocorrelation reduces the amount of information available about $\bm\pi$ (cf. Section~\ref{s.autocorr}). 
The standard error $\text{SE}(\hat\pi_i)=\sqrt{\hat\pi_i (1-\hat\pi_i)/T}$ that assumes independent sampling will obviously underestimate the true variability of the estimate $\bm{\hat\pi}$ if samples are correlated \citep[][]{green1995reversible, sisson2005transdimensional}. 
To obtain a measure of precision, \citet{green1995reversible} proposed running several independent MCMC chains $c=1,\dots,C$ and computing \cc the standard deviation of the estimates $\bm{\hat\pi}^{(c)}$ \sc across these independent replications. 
\cc However, for complex models, this method might require a substantial amount of additional computing time for burn-in and adaption and thus can be infeasible in practice. \sc

Assessing the precision of the estimate $\bm{\hat\pi}$, which depends on the autocorrelation of the sequence of discrete \cc MCMC \sc samples $z^{(t)}$, is of major importance. 
In case of model selection, it must be ensured that the estimated posterior probabilities $p(\mathcal M_i \mid \bm y)$ are sufficiently precise for drawing substantive conclusions. 
This issue is especially important when estimating a ratio of marginal probabilities, that is, the Bayes factor $B_{ij}=p(\bm y\mid \mathcal M_i)/p(\bm y\mid \mathcal M_j)$ \citep{jeffreys1961theory}. 
Moreover, it is often of interest to compute the effective sample size defined as the number of independent samples that would provide the same amount of information as the given MCMC output for estimating $\bm\pi$ with $\bm{\hat\pi}$. 
Besides providing an intuitive measure of precision, a minimum effective sample size can serve as a principled and theoretically justified stopping rule for MCMC sampling \citep{gong2016practical}. 
However, standard methods of estimating the effective sample size \citep[e.g., computing the spectral density at zero;][]{plummer2006coda,heidelberger1981spectral} are tailored to continuous parameters.
When applied to the model-indexing variable $z^{(t)}$ of a transdimensional MCMC method, these methods neglect the discreteness of $z^{(t)}$.
Depending on the specific numerical labels used for the different models (e.g., $(1,2,3,4)$ vs. $(1,4,2,3)$), spectral decomposition can lead to widely varying and arbitrary estimates for the effective sample size (see Section~\ref{s.logistic}).

In summary, transdimensional MCMC is a very important and popular method for Bayesian inference \citep{sisson2005transdimensional}. 
However, little attention has been paid to the analysis of the resulting MCMC output, which requires that one takes into account the autocorrelation as well as the discrete nature of the model-indexing variable. 
As a solution, we propose to fit a discrete, first order Markov model to the MCMC output $z^{(t)}$ to assess the precision of the estimated stationary distribution $\bm{\hat\pi}$. 
Whereas several diagnostics have previously been proposed to assess the \textit{convergence} of transdimensional MCMC samplers \citep[e.g.,][]{brooks2000markov, castelloe2002convergence, brooks2003nonparametric, sisson2007distancebased}, we are unaware of any methods that quantify the \textit{precision} of the point estimate $\bm{\hat\pi}$.

\section{Method}

\subsection{A Discrete Markov Model for Transdimensional MCMC Output}
\label{s.markov}

The proposed method approximates the output of a transdimensional MCMC method (i.e., the sampled iterations $z^{(t)}$) by a discrete Markov model $\mathcal M^\text{Markov}$ with transition matrix $\bm P$.
This model explicitly accounts for autocorrelation, which in turn allows quantifying estimation uncertainty for the discrete stationary distribution $\bm\pi$.  
The entries of $\bm P$ are defined as the transition probabilities $p_{ij} = P(z^{(t+1)}=j \mid z^{(t)}=i)$ for all $i,j=1,\dots,I$, with rows summing to one, $\sum_{j=1}^I p_{ij} = 1$. 
According to the discrete Markov model, the probability distribution of the indexing variable $z^{(t)}$ at iteration $t$ is given by multiplying the transposed initial distribution $\bm \pi_0^\top$ by the transition matrix $t$ times, $P(z^{(t)}=i)= [\bm \pi_0^\top \bm P^t]_i$. 
The proposed method estimates the transition matrix $\bm P$ as a free parameter based on the sufficient statistic $\bm N$, the matrix of frequencies $n_{ij}$ counting the observed transitions from $z^{(t)}=i$ to $z^{(t+1)}=j$  \citep{anderson1957statistical}.

Due to the construction of the transdimensional MCMC sampler, the discrete indexing variable $z^{(t)}$ follows a stationary distribution with a constant probability vector $\bm \pi$ (i.e., the posterior model probabilities of interest). 
Hence, when modeling the sequence $z^{(t)}$ with the discrete Markov model $\mathcal M^\text{Markov}$, this implies that the transition matrix $\bm P$ must satisfy the condition for stationarity
\begin{equation}
\label{e.eigen}
\bm \pi^\top \bm P = 1 \cdot \bm \pi^\top,
\end{equation}
meaning that the probability vector $\bm \pi$ is the left eigenvector of the matrix $\bm P$ with eigenvalue one \citep[with $\bm \pi$ normalized to sum to one;][]{anderson1957statistical}. 
Based on the model $\mathcal M^\text{Markov}$, an estimator for $\bm \pi$ is thus obtained by computing the eigenvector of $\bm P$ with eigenvalue one \citep{barker2013bayesian}. 

However, we are less interested in a new estimator $\bm{\hat\pi}$ of the stationary distribution but rather in the precision of this estimate. 
To quantify estimation uncertainty, we thus fit the model $\mathcal M^\text{Markov}$ with $\bm P$ as a free parameter in a Bayesian framework by drawing posterior samples $\bm P^{(r)}$ ($r=1,\dots,R$ ).
\cc Similar to a parametric bootstrap, \sc this Bayesian sampling approach has the advantage that we can easily quantify estimation uncertainty (i.e., the dispersion of the posterior distribution of $\bm P$) by computing descriptive statistics of the samples $\bm P^{(r)}$ (e.g., the standard deviation or credibility intervals).
Moreover, we can directly quantify the estimation uncertainty of derived quantities such as the posterior model probabilities, model ranks, or Bayes factors (see Section~\ref{s.precision}).
In the following, it is important to distinguish between the posterior distribution of $\bm P$ given the sufficient statistic $\bm N$, which quantifies the uncertainty of $\bm P$ due to estimation error of the transdimensional MCMC method, and \cc the posterior distribution of the models given the empirical data, which is represented by the constant vector of probabilities $\bm\pi$ for a specific data set. \sc 

Next, we define a prior distribution for the parameter $\bm P$ of the model $\mathcal M^\text{Markov}$.
Given that the transition matrix $\bm P$ includes one probability vector $\bm p_{i}$ for each row $i$, we assume independent Dirichlet distributions with parameter $\epsilon\geq 0$ for each row, 
\begin{equation}
\bm p_{i} \equiv (p_{i1},\dots,p_{iI}) \sim \mathcal{D}(\epsilon,\dots , \epsilon).
\label{e.prior}
\end{equation} 
Conditional on the MCMC output $\bm N$, the estimation uncertainty of $\bm P$ is approximated by drawing $R$ posterior samples $\bm P^{(r)}$. 
Since the Dirichlet prior is conjugate to the multinomial distribution, independent samples $\bm P^{(r)}$ can efficiently be drawn from the Dirichlet distribution with parameters
\begin{equation}
\label{e.post}
\bm p_{i}^{(r)} \sim \mathcal{D}(n_{i1} + \epsilon,\dots , n_{iI} + \epsilon).
\end{equation} 
Based on these samples, the estimation uncertainty of the stationary probabilities $\bm \pi$ is assessed by computing the (normalized) eigenvector with eigenvalue one for each sample $\bm P^{(r)}$ (Eq.~\ref{e.eigen}).
\cc Algorithm~\ref{a.precision} provides an overview of the computational steps of the proposed method as pseudo-code. \sc

\begin{algorithm}[t] \setstretch{1.5} 
\caption{\cc Quantify uncertainty of $\hat{\bm\pi}$ due to transdimensional MCMC sampling. \cc}   \label{a.precision} 
\begin{algorithmic}[1] 
\Procedure{Markov Model}{}
\State Sampling $z^{(t)}$: $T$ iterations of model-indexing variable $z$ via transdimensional MCMC
\State Compute $\bm N$: Observed $I \times I$ transition matrix of $z ^{(t)}$ with elements $n_{ij}$
\State Set prior parameter $\epsilon$ 
(default: $\epsilon = 1/I^*$ for the $I^*$ models observed in $z^{(t)}$, $\epsilon = 0$ otherwise)
\For { $r = 1,\dots,R$ }
	\State Initialize posterior sample $\bm P^{(r)}$: $I \times I$ transition matrix  with rows $\bm p_i^{(r)}$
	\For { $i = 1,\dots,I$ }
		\State Sampling $\bm p_i^{(r)} \sim$ Dirichlet($n_{i1} + \epsilon,\dots , n_{iI} + \epsilon$)
	\EndFor
	\State Initialize posterior sample $\bm \pi^{(r)}$: Posterior model probabilities 
	\State $\bm \pi^{(r)} \gets$ (normalized) eigenvector of $\bm P^{(r)}$ with eigenvalue one
\EndFor
\If {(quantify uncertainty)}
\State Compute summary statistic for all samples $\bm\pi^{(r)}$ 
\State Example: $\text{SD}_\text{Markov}(\hat\pi_i) \gets \text{SD}(\pi_i^{(r)})$
\EndIf
\If {(compute effective sample size)}
\State Using all $\bm\pi^{(r)}$: Fit Dirichlet parameters $\hat\alpha_1,\dots,\hat\alpha_I$  \citep[][]{minka2000estimating}
\State Compute effect sample size $\widehat T_\text{eff}
\gets \sum_{i=1}^I \widehat\alpha_i - (I^*)^2 \epsilon$
\EndIf
\EndProcedure
\end{algorithmic}
\end{algorithm} 
\sc

With regard to the prior parameter $\epsilon$, small values should be chosen to reduce its influence on the estimation of $\bm P$.
In principle, the improper prior $\epsilon = 0$ can be used, which minimizes the impact of the prior on the estimated stationary distribution. 
This improper prior also ensures that the results do not hinge on the set of models that could possibly be sampled, but were never actually observed in the sequence $z^{(t)}$. 
For such unsampled models, the corresponding rows and columns of the observed transition matrix $\bm N$ are filled with zeros. 
With $\epsilon=0$, the relevant eigenvector of the transition matrix $\bm P \mid \bm N$ is thus identical to that of a reduced matrix $\bm {P^*} \mid \bm {N^*}$ that includes only the transitions for the subset of models sampled in $z^{(t)}$.
However, in our simulations, this improper Dirichlet prior proved to be numerically unstable and resulted in more variable point estimates than the standard i.i.d. estimate or the proper prior discussed next. 

Here, we use the weakly informative prior $\epsilon = 1/I$ as a default, which has an impact equivalent to one observation for each row of the observed transition matrix $\bm N$. 
By putting a small weight on all values of the transition matrix $\bm P$, this prior serves as a regularization of the posterior \citep{alvares2018what}.
However, in scenarios where the number of models exceeds the number of iterations of the transdimensional MCMC method (i.e., $I\gg T$), such a regularization assigns substantial probability weight to models that are never observed in $z^{(t)}$.
To limit the effect of the prior, we thus set $\epsilon = 1/I^*$ only for those $I^*$ models that were observed in $z^{(t)}$ and $\epsilon = 0$ for the remaining models. 
Besides reducing the impact of the prior, this choice has the computational advantage that \cc one can draw posterior samples and compute eigenvectors for the reduced matrix $\bm {P^*} \mid \bm {N^*}$ that includes only the sampled models. \sc
In the two examples in Sections~\ref{s.logistic} and \ref{s.heart}, this prior has proved to be numerically robust and resulted in point estimates close to the standard i.i.d. estimates.

As a third alternative, the prior can be adapted to the structure of specific transdimensional MCMC implementations, which only implement switches to a small subset of the competing models. 
For instance, in variable selection, regression parameters are often added or removed one at a time, resulting in a birth-death process \citep{stephens2000bayesian}. 
For these kinds of samplers, the Dirichlet parameters $\epsilon_{ij}$ can be set to zero selectively. 
However, such adjustments will be dependent on the chosen MCMC sampling scheme. 
The default choice of $\epsilon = 1/I^*$ for sampled models and $\epsilon = 0$ for nonsampled models 
provides a good compromise of being very general and numerically robust, while having a small effect on the posterior.
However, in general, the choice of $\epsilon$ becomes less influential as the number of MCMC samples increases (especially if the row sums of $\bm N$ are large).

\subsection{\cc Estimation Uncertainty}
\label{s.precision}

Based on the \cc posterior samples $\bm P^{(r)}$ of the transition matrix and the derived model probabilities $\bm \pi^{(r)}$, \sc it is straightforward to estimate the stationary distribution by the posterior mean $\bm{\hat \pi}$ (alternatively, the median or mode may be used). 
More importantly, however, estimation uncertainty due to the transdimensional MCMC method can directly be assessed by plotting the estimated posterior densities for each $\pi_i$. 
To quantify the precision of the estimate $\bm{ \hat\pi}$, one can report posterior standard deviations or credibility intervals for the components $\hat\pi_i$. 
These component-wise summary statistics are most useful if the number of models $I$ is relatively small. 

An important advantage of drawing posterior samples $\bm \pi^{(r)}$ in a Bayesian framework (instead of using asymptotic approximations for the standard error of $\hat{\bm\pi}$) is that one can directly quantify estimation uncertainty for other quantities of interest.
For very large numbers of sampled models, the assessment of estimation uncertainty can be focused on the subset of $k$ models with the highest posterior model probabilities. 
Within the sampling approach, estimation uncertainty for the $k$ best-performing models can easily be assessed by computing ranks for each of the posterior samples $\bm \pi^{(r)}$. 
Then, the variability of these model ranks across the $R$ samples can be summarized, for instance, by the percentage of identical rank orders for the $k$ best-performing models, or the percentages of how often each model is included within the subset of the $k$ best-performing models (i.e., has a rank smaller or equal to $k$). 

In case of model selection, dispersion statistics such as the posterior standard deviation are also of interest with respect to the Bayes factor $B_{ij}$ \citep{kass1995bayes}. 
To judge the estimation uncertainty for the Bayes factor, one can evaluate the corresponding posterior distribution by computing the derived quantities $B_{ij}^{(r)} = \pi_i^{(r)}/\pi_j^{(r)}$ (given uniform prior model probabilities). 
Precision can also be assessed for model-averaging contexts when comparing subsets of models against each other (e.g., regression models including a specific effect vs. those not including it). 
Given such disjoint sets of model indices $M_s\subset \{1,\dots,I\}$, the posterior probability for each subset of models is directly obtained by summing the posterior samples $\pi_i^{(r)}$ for all $i \in M_s$. 
\cc Note that it is invalid to aggregate across model subsets or to drop sampled models before applying the proposed Markov approach because functions of discrete Markov chains (e.g., collapsing the $I$ original states into a subset of $S$ states) are not Markovian in general \citep{burke1958markovian}. \sc

\subsection{Effective Sample Size}
\label{s.neff}

Besides quantifying estimation uncertainty, the posterior samples $\bm\pi^{(r)}$ can be used to estimate the effective sample size for the transdimensional MCMC output. 
For this purpose, we consider the benchmark model $\mathcal M^\text{iid}$ under the ideal scenario of drawing independent samples $\tilde z^{(t)}$ from the categorical distribution with probabilities $\bm {\tilde \pi}$. 
For this model, we assume an improper Dirichlet prior on the stationary distribution, $\bm {\tilde \pi} \sim \mathcal D(0,\dots, 0)$ (whereas the Markov model assumes a Dirichlet prior on the transition probabilities).
Since this prior is conjugate to the multinomial distribution, the posterior for the stationary distribution $\bm {\tilde \pi}$ is given by 
\begin{equation}
\label{e.postiid}
\bm {\tilde\pi} \mid \bm{\tilde\bm n} \sim \mathcal D (\tilde n_1, \dots, \tilde n_I),
\end{equation}
conditional on the observed frequencies $\cc {\tilde n_i}=\sum_{t=1}^T \mathbb I(\tilde{z}^{(t)} = i)$. 
Note that the transition frequencies are rendered irrelevant in this i.i.d. model, \cc since there are no dependencies \sc in the sampled iterations $\tilde{z}^{(t)}$.

Given the \cc dependent samples \sc $z^{(t)}$ of a transdimensional MCMC chain, we can now compare the empirical posterior distribution of $\bm\pi$ estimated with the model $\mathcal M^\text{Markov}$ against the theoretically expected posterior distribution of $\bm {\tilde \pi}$ under the hypothetical model $\mathcal M^\text{iid}$. 
Essentially, we match the latter distribution to the former to estimate the effective sample size as the total number of independent samples \cc $T_\text{iid} = \sum_i \tilde n_i$ \sc that would result in a similar dispersion as that estimated by the Markov model.
\cc To estimate the $\tilde n_i$, the i.i.d. posterior distribution in Eq.~\ref{e.postiid} is fitted to the posterior distribution of the Markov model by estimating the shape parameters $\alpha_1,\dots,\alpha_I$ of a Dirichlet distribution given the sampled $\bm\pi^{(r)}$ \citep[which can be achieved by an efficient maximum-likelihood algorithm by][see Appendix]{minka2000estimating}. \sc 
Next, a comparison of the estimated Dirichlet parameters $\widehat \alpha_i$ with the conjugate posterior in Eq.~\ref{e.postiid} yields \cc $\widehat{\tilde n} = \widehat \alpha_i$, which implies that the dispersion of the posterior model probabilities $\bm\pi^{(r)}$ is equivalent to having observed $\widehat T_\text{iid} = \sum_i \hat \alpha_i$ independent samples.
However, the samples $\bm\pi^{(r)}$ are not only informed by the samples $z^{(t)}$ of the transdimensional MCMC sampler, but also by the prior distribution of the Markov model, which is irrelevant for estimating the effective sample size.
Hence, to estimate the effective sample size for the transdimensional MCMC sampler, it is necessary to subtract the prior sample size $I^2\epsilon$ of the Markov model (cf. Eq.~\ref{e.prior}), which reflects the relative weight of the prior, since the Dirichlet shape parameter $\epsilon$ occurs $I$ times in each row of the $I\times I$ transition matrix $\bm P$ \citep{alvares2018what}. \sc
Overall, it follows that the effective sample size under the assumption of independent sampling from a multinomial distribution is estimated as
\begin{equation} \label{e.neff}
\widehat T_\text{eff} = \sum_{i=1}^I \widehat\alpha_i - I^2 \epsilon.
\end{equation}
\cc Note that it is necessary to replace $I$ by $I^*$ in Eq.~\ref{e.neff} if the Markov model uses only those $I^*$ models that were actually sampled in $z^{(t)}$. \sc
Importantly, the estimate $\widehat T_\text{eff}$ takes the discreteness of the indexing variable $z$ into account and \cc does not depend on specific (but arbitrary) numerical values of the model indices. \sc

\subsection{Remarks}

The proposed method quantifies estimation uncertainty by fitting a discrete Markov model to transdimensional MCMC output. 
For this purpose, a simplifying assumption is made that is not guaranteed to hold. 
Whereas samples of the full model space $(z^{(t)},\bm\theta^{(t)})$ necessarily follow a Markov process by construction, this does not imply that the samples $z^{(t)}$ follow a Markov chain marginally \citep{brooks2003efficient, lodewyckx2011tutorial}. 
In practice, the iterations of the model-indexing variable $z^{(t)}$ might have higher-order dependencies since transition probabilities depend on the exact state of the MCMC sampler in each of the models' parameter spaces.
However, in Sections~\ref{s.logistic} and \ref{s.heart} we show in two empirical examples that the proposed simplification (i.e., fitting a Markov chain of order one) is sufficient to account for autocorrelations in the samples $z^{(t)}$ in practice.
Moreover, the approximation by a first-order Markov chain provides a trade-off between ignoring dependencies completely (i.e., assuming i.i.d. samples) and accounting for any higher-order dependencies (which will likely increase the computational burden especially for large numbers of models).
Note that it is a common practice to rely on simplifying assumptions for the analysis of simulation output; for instance, a standard approach of estimating the effective sample size for continuous parameters assumes that the output sequence can be modeled as a covariance stationary process with a smooth log spectrum \citep[][]{heidelberger1981spectral}.

The proposed method of fitting a discrete Markov model is very general and can be applied irrespective of specific transdimensional MCMC implementations.
Moreover, it requires only the sampled sequence $z^\text{(t)}$ of the discrete parameter or the  matrix $\bm N$ with the observed frequency of transitions. 
\cc If output from multiple independent chains $c=1,\dots,C$ is available, the transition frequency matrices $\bm N^{(1)},\dots, \bm N^{(C)}$ can simply be summed before applying the method. \sc 
This follows directly from Bayesian updating of the stationary distribution $\bm \pi$. 
Essentially, each chain provides independent evidence for the posterior of the transition matrix $\bm P$, which is reflected by using the sums $\sum_c n_{ij}^{(c)}$ for the conjugate Dirichlet prior in Eq.~\ref{e.post}.
Note that this feature can be used to compare the efficiency of many short versus few long MCMC chains.

In the R package \texttt{MCMCprecision} \citep[][]{heck2018mcmcprecision}, we provide an implementation that relies on the efficient computation of eigenvectors in the C++ library \texttt{Armadillo} \citep{sanderson2016armadillo}, accessible in R via the package \texttt{RcppArmadillo} \citep{eddelbuettel2014rcpparmadillo}. 
On a notebook with an Intel\textsuperscript{\textregistered} i7-7700HQ processing unit, drawing $R=5,000$ samples from the posterior distribution for 10 (100) sampled models requires approximately 150 milliseconds (28 seconds). 
\cc Similar to any MCMC or bootstrap approach, the choice of the number of samples $R$ depends on the summary statistic used to quantify uncertainty. 
Whereas more samples are required to approximate the density distribution (e.g.,  $R \geq 5,000$), less samples (e.g.,  $R \approx 1,000$) are sufficient to approximate the SD of the estimated posterior model probabilities. 
Since the samples $\bm\pi^{(r)}$ are independently drawn and SDs are usually sufficient to quantify uncertainty, the choice $R=1,000$ is often sufficient in practice (however, for the simulations below, we use $R=5,000$).
\sc

\section{Illustration: Effect of Autocorrelation}
\label{s.autocorr}

Before applying the proposed method to actual output of transdimensional MCMC samplers, we first illustrate its use in an idealized setting, \cc where the interest is in approximating the posterior model probabilities $\bm\pi=(.85 , .13 ,.02)^\top$ by drawing random samples $z^{(t)}$. \sc 
To investigate the effect of \cc independent versus dependent sampling, \sc we generated sequences $z^{(t)}$ from the Markov model $\mathcal M^\text{Markov}$ with the stationary distribution $\bm\pi$.
To induce autocorrelation, we defined a mixture process for each iteration $t$. 
With probability $\beta$, the discrete indexing variable was identical to the current model, $z_{t+1}=z_t$. 
In contrast, with probability $1-\beta$, the value $z_{t+1}$ was sampled from the given stationary distribution $\bm \pi$. 
Thereby, increasing values of $\beta$ resulted in a larger autocorrelation of the sequence $z^{(t)}$ \cc as illustrated for $\beta=0$ and $\beta=0.8$ in Figure~\ref{f.chain}A and \ref{f.chain}B, respectively. \sc

For varying levels of $\beta=0,0.1,\dots,0.8$, we sampled 500 replications with $T=1,000$ iterations each, applied the proposed method (with $R = 5,000$) and computed the precision of the estimate $\bm{\hat\pi}$.
\cc The main interest is in the posterior SD and in the coverage probability, defined as the probability that the data-generating values $\bm\pi$ are in the 90\% credibility interval defined by the 5\% and 95\% quantiles.  \sc 
As a benchmark, we also computed these summary statistics under the (false) assumption that the samples $z^{(t)}$ were independently drawn by \cc fitting the model $\mathcal M^\text{iid}$  with the Dirichlet posterior distribution in Eq.~\ref{e.postiid}. \sc 
Note that the latter uncertainty estimate is equivalent to the standard Monte Carlo error that assumes independent sampling.

\begin{figure}[!tbh]
\includegraphics[width=16cm]{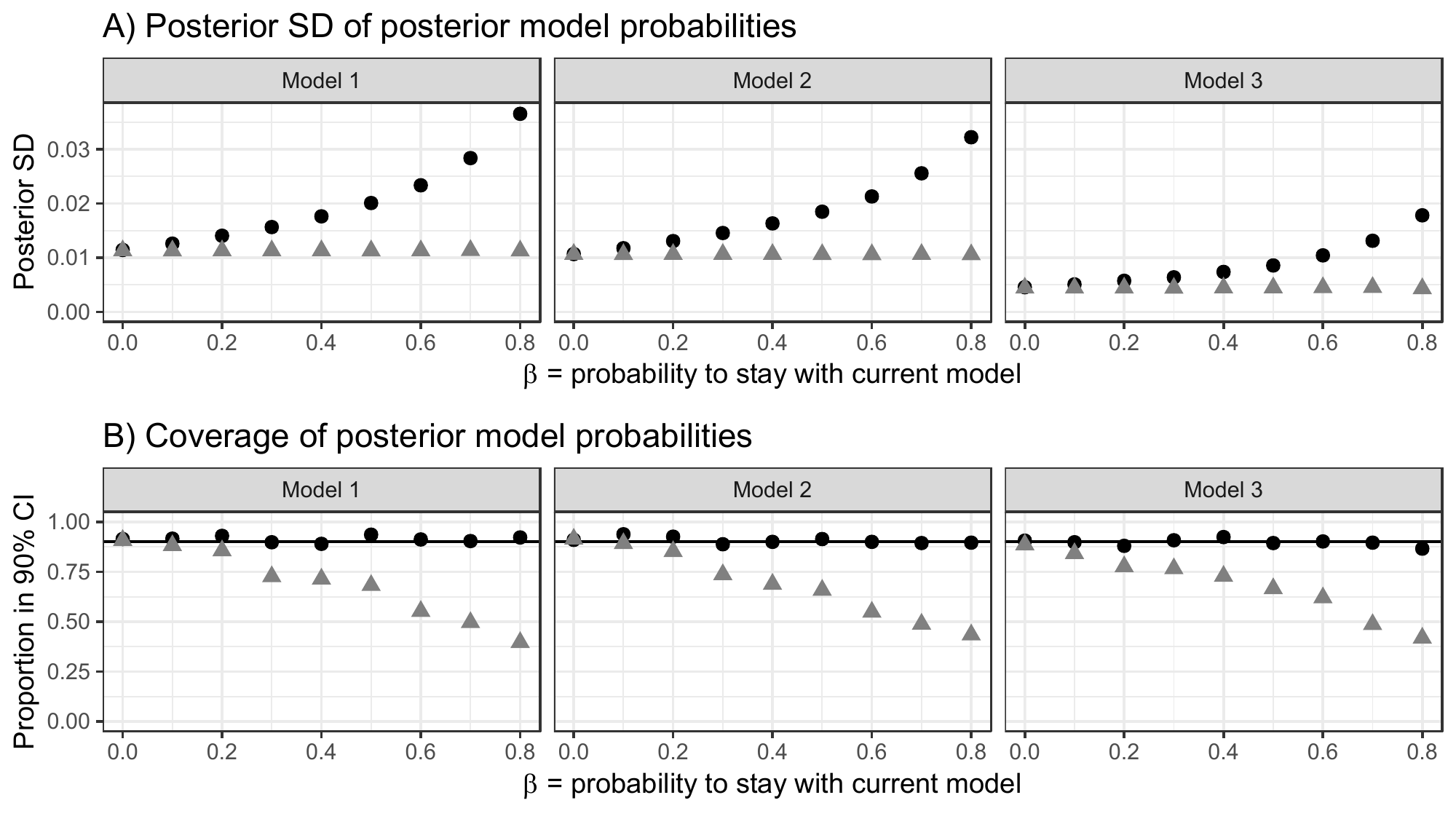}
\caption{
Estimation uncertainty for the stationary distribution $\bm \pi$. 
(A) The Markov method (black dots) correctly indicated that estimation error of the posterior model probabilities increased as autocorrelation increased. 
When assuming i.i.d. sampling (gray triangles), the estimated precision did not depend on the autocorrelation. 
(B) Proportion of 500 replications for which the 90\% CI intervals included the data-generating stationary distribution $\bm\pi$.
}
\label{f.illu}
\end{figure}

Figure~\ref{f.illu} shows the results of this simulation. 
In Figure~\ref{f.illu}A, \cc the three panels correspond to the estimation uncertainty (i.e., the posterior SD) of the three posterior model probabilities $\bm\pi=(\pi_1,\pi_2,\pi_3)^\top$.
The estimated posterior SD of the Markov model indicated increasing uncertainty for larger values of $\beta$, thus taking the increasing autocorrelation into account.
In contrast, \sc the model $\mathcal M^\text{iid}$ assumes independence a priori, and thus, the posterior uncertainty was independent of $\beta$. 
As a result of this, the corresponding 90\% credibility interval was less likely to include the data-generating value $\bm\pi$ for increasing values of $\beta$ (see Figure~\ref{f.illu}B), whereas the Markov model provided an accurate description of the estimation uncertainty \cc for any degree of dependence. \sc

\section{Variable Selection in Logistic Regression}
\label{s.logistic}

In the following, we apply the proposed method to the problem of selecting variables in a logistic regression, an example introduced by \citet{dellaportas2000bayesian} to highlight the implementation of transdimensional MCMC in \texttt{BUGS} \citep[see also][]{dellaportas2002bayesian,ntzoufras2002gibbs}. 
Table~\ref{t.healy} shows the frequencies of deaths and survivals conditional on severity and whether patients received treatment \citep[i.e., antitoxin medication;][]{healy1988glim}. 
To emphasize the importance of considering estimation uncertainty for the posterior model probabilities, we compare the efficiency of two transdimensional MCMC approaches, which can both be implemented in \texttt{JAGS} \citep{plummer2003jags}.

\begin{table}
\centering
\caption{Logistic regression data set by \citet{healy1988glim}.}
\label{t.healy}
\begin{tabular}{ccrr}
\midrule
Condition (A)  & Antitoxin (B)   & Death & Survival \\ 
\midrule
More~Severe &Yes& 15&6\\
 			&No & 22 & 4 \\
Less~Severe &Yes& 5 & 15\\
		    &No & 7 & 5 \\ 
\midrule
\end{tabular}
\end{table}

The full logistic regression model assumes a \cc binomial distribution $\mathcal {B}$ \sc of the survival frequencies $y_{jl}$ and a linear model on the logit-transformed survival probabilities $p_{jl}$,
\begin{align}
y_{jl} &\sim \mathcal {B}(p_{jl}, n_{jl})\\
\log\left(\frac{p_{jl}}{1-p_{jl}}\right) &= \beta_0 + \beta_1 a_j+ \beta_2 b_l+ \beta_3 (ab)_{jl},\hspace{.5cm} j,l=1,2
\label{e.logistic}
\end{align}
where $n_{jl}$ are the total number of patients in condition $jl$ and $\bm\beta$ the regression coefficient for the effect-coded variables $a_{j}$, $b_{l}$, and $(ab)_{jl}$. 
Variable selection is required to choose between $I=5$ models: 
the intercept-only model I, the three main effect models A, B, and A+B, and the model AB that includes the interaction. 
For comparability, we use the same priors as  \citet{dellaportas2000bayesian} and assume a centered Gaussian prior with variance $\sigma^2=8$ for each regression parameter, $\beta_k \sim \mathcal N(0,8)$. 
Moreover, the model probabilities were set to be uniform, $p(\mathcal M_i)=1/5$. 
Note that nonuniform prior probabilities might be used to protect against multiple testing issues \citep[i.e., Bayes multiplicity;][]{scott2010bayes}.

One of the two implemented transdimensional MCMC approaches uses unconditional priors  \citep[][KM98]{kuo1998variable} and includes indicator variables $\gamma_{ik} \in \{0,1\}$ for each regression coefficient $\beta_k$ in model $\mathcal M_i$. 
The parameter $\bm \gamma_i$ determines which regression coefficients are included by removing some of the additive terms of the linear model in Equation~\ref{e.logistic}. 
Details about the full and conditional posterior distributions are provided by \citet[][p. 7]{dellaportas2000bayesian}.

As a second transdimensional MCMC approach, we implemented the method of Carlin and Chib (\citeyear{carlin1995bayesian}; CC95), which stacks up all model parameters into a new parameter $\bm\theta= (z,\bm\beta_1,\dots, \bm\beta_I)$, where $\bm\beta_i$ is the vector of regression parameters of model $\mathcal M_i$. 
Thereby, this approach samples a total of 12 regression parameters along with the indexing variable $z$.  
Note that the method of \citet{carlin1995bayesian} uses pseudo-priors $p(\bm\beta_{i} \mid \mathcal M_{j})$, $i \neq j$, that do not influence the statistical inference about $p(\bm y \mid \mathcal M_i)$ and $p(\bm\beta_i \mid \bm y, \mathcal M_i)$. 
However, these pseudo-priors  determine the conditional proposal probabilities $p(z\mid \bm y,  \bm\beta_1,\dots, \bm\beta_I)$ of switching between the models and thereby affect the efficiency of the MCMC chain. 
In substantive applications, these pseudo-priors should be chosen to match the posterior $p(\bm\beta_{i} \mid \mathcal M_{i})$ in order to ensure high probabilities of switching between the models \citep[cf.][]{carlin1995bayesian,barker2013bayesian}. 
Here, however, \cc we did not optimize the sampling scheme and used $\bm\beta_{ik} \mid \mathcal M_{j}\sim \mathcal N(0,8)$ for the pseudo-priors to illustrate that our method can correctly detect the lower precision resulting from this suboptimal choice. \sc 

\begin{figure}[th]
\includegraphics[width=16cm]{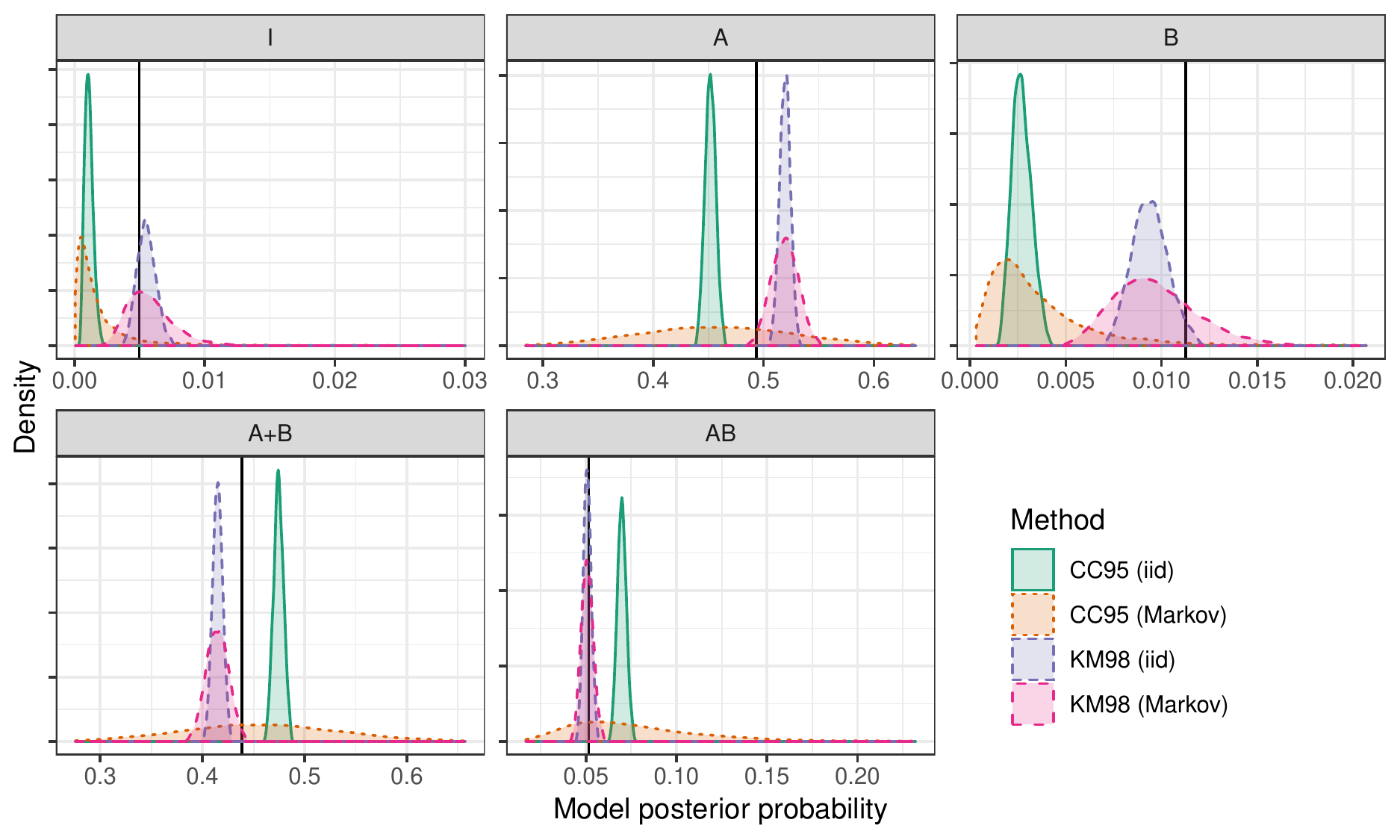}
\caption{
\cc The five panels show the estimation uncertainty of the posterior model probabilities $\bm\pi=(\pi_1,\dots,\pi_5)^\top$ for the five logistic regression models I (intercept only), A, B, and A+B (only main effects), and AB (two main effects and interaction).
For both transdimensional MCMC samplers (CC95 = \citealp{carlin1995bayesian}; KM98 = \citealp{kuo1998variable}), the posterior distribution of the Markov model included the correct reference values (vertical black lines) with high probability.
In contrast, the i.i.d. model underestimated estimation uncertainty and posterior distributions did not include the target values with high probability. \sc
}
\label{f.posterior}
\end{figure}

Figure~\ref{f.posterior} shows the estimated posterior distribution \cc ($R = 5,000$) \sc of the posterior model probabilities using one Markov chain with 11,000 iterations (including 1,000 burn-in samples). 
The vertical black lines show the \cc reference values for $\bm\pi$, approximated with very high accuracy by the KM98 approach using eight independent chains and one million samples each. \sc 
As expected, the (incorrect) assumption that $z^{(t)}$ are sampled independently resulted in overconfidence in the point estimates of the CC95 approach. 
For all models, the corresponding posterior distributions missed the correct value and did not identify this estimation uncertainty.
\cc This shows the importance of assessing the dependency in the samples $z^{(t)}$ in order to judge the estimation uncertainty for the estimated posterior model probabilities. 
As a remedy, \sc the proposed Markov approach resulted in a posterior distribution that covered the target values with high probability.
Moreover, \cc  the novel estimation method revealed that the KM98 implementation had a higher precision compared to the CC95 approach, which was likely due to the (intentionally not optimized) choice of the pseudo-priors in the latter method. 
Hence, the Markov model allows comparison of the estimation uncertainty of different transdimensional MCMC methods for the model probabilities $\bm\pi$. \sc

\begin{table}[ht]
\centering
\caption{Estimated posterior model probabilities in percent.}
\label{t.logistic}
\begin{threeparttable}
\begin{tabular}{crrrrrrrr}
  \hline
  &\multicolumn{4}{c}{\citet{kuo1998variable}} & 
   \multicolumn{4}{c}{\citet{carlin1995bayesian}} \\ 
  \cmidrule(lr){2-5}\cmidrule(lr){6-9} 
Model &   
  $\text{Mean}(\bm{\hat\pi})$ 
& $\text{SD}  (\bm{\hat\pi})$ 
& \cc $\overline{\text{SD}}_\text{iid}$ 
& \cc $\overline{\text{SD}}_\text{Markov}$   
& $\text{Mean}(\bm{\hat\pi})$  
& $\text{SD}  (\bm{\hat\pi})$
& \cc $\overline{\text{SD}}_\text{iid}$ 
& \cc $\overline{\text{SD}}_\text{Markov}$ \\ 
  \hline
  1 & 0.51 & 0.24 & 0.07 & 0.16 & 0.57 & 0.35 & 0.06 & 0.39 \\ 
  A & 49.28 & 1.38 & 0.50 & 1.22 & 48.55 & 7.14 & 0.49 & 6.92 \\ 
  B & 1.14 & 0.44 & 0.10 & 0.26 & 1.26 & 0.63 & 0.10 & 0.73 \\ 
  A+B & 43.85 & 1.25 & 0.50 & 1.10 & 43.61 & 7.41 & 0.49 & 7.19 \\ 
  AB & 5.22 & 0.37 & 0.22 & 0.34 & 6.00 & 3.38 & 0.21 & 3.82 \\ 
\hline
\end{tabular}
\begin{tablenotes}
\footnotesize\note
Posterior model probability estimates $\bm{\hat\pi}$ are shown in percent. 
$\text{Mean}(\bm{\hat\pi})$ and $\text{SD}(\bm{\hat\pi})$ were computed across \cc 500 \sc replications.  
As a measure for the estimated precision, means of the posterior SD are shown ($\cc\overline{\text{SD}}_\text{iid}$ assumes independent sampling; $\cc\overline{\text{SD}}_\text{Markov}$ assumes a Markov chain model).
\end{tablenotes}
\end{threeparttable}
\end{table}

To test the validity of the proposed method more rigorously, we replicated the previous analysis 500 times. 
Thereby, the estimated precision can be compared against the actual sampling variability of the estimated model probabilities. 
For both transdimensional MCMC methods, Table~\ref{t.logistic} shows the mean estimated model probabilities in percent. 
Across replications, the point estimates (posterior means) from the Markov and the i.i.d. approach were very similar with a median absolute difference of $0.03\%$ and $0.31\%$ for the KM98 and CC95 implementations, respectively. 
To judge whether the estimated precision (i.e., the mean posterior standard deviations $\cc\overline{\text{SD}}_\text{iid}$ and $\cc\overline{\text{SD}}_\text{Markov}$) is valid, Table~\ref{t.logistic} shows the empirical SD of the estimates $\bm{\hat\pi}$ 
across replications. 
The results show that the assumption of independent samples $z^{(t)}$ leads to an overconfident assessment of the precision for the estimated model probabilities,  $\overline{\text{SD}}_\text{iid} \ll \text{SD}(\bm{\hat\pi})$, which is especially severe for the less efficient CC95 implementation.
In contrast, the Markov approach provided good estimates of the actual estimation uncertainty, $\overline{\text{SD}}_\text{Markov} \approx \text{SD}(\bm{\hat\pi})$. 
Moreover, for the MCMC method by \citet{carlin1995bayesian}, the larger SDs indicate a smaller efficiency compared to the unconditional prior approach by \citet{kuo1998variable}. 
\cc This theoretically expected result is due to the suboptimal choice of pseudo-priors. \sc 
However, note that this difference in efficiency may be overlooked when merely computing relative proportions based on the sampled indexing variable $z^{(t)}$ (i.e., when implicitly assuming independent samples).

\begin{figure}
\includegraphics[width=12cm]{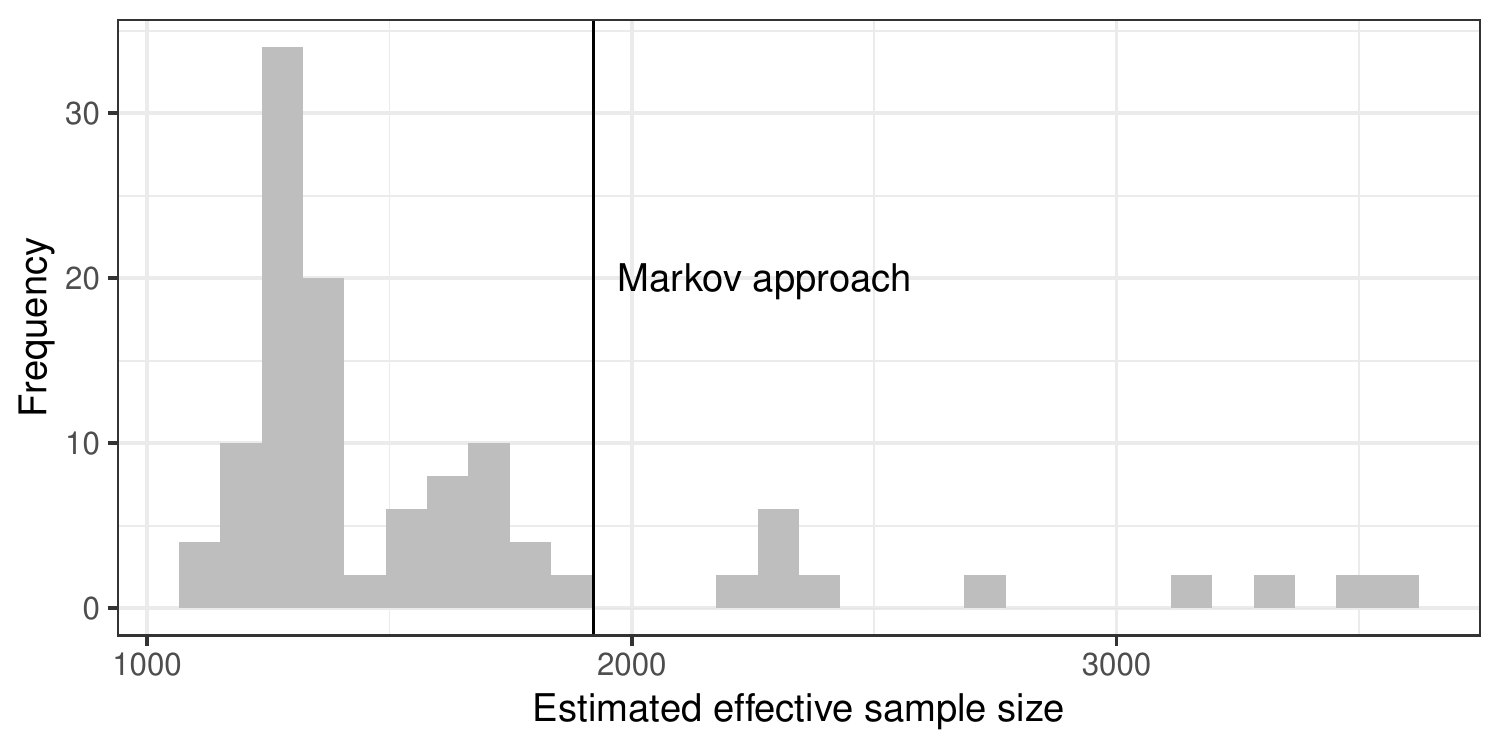}
\caption{Effective sample size as estimated by the spectral density at zero \citep{plummer2006coda} for all permutations of \cc the model indices for a given MCMC output $z^{(t)}$ \sc (based on 10,000 samples of the method by \citealp{kuo1998variable}).}
\label{f.neff}
\end{figure}

The higher efficiency of the KM98 approach becomes even clearer when assessing the median of the estimated effective sample size, which was $2,043$ for the KM98 approach compared to only $65$ for the CC95 method. 
As discussed above, commonly used estimators of effective sample size for continuous parameters \citep[e.g.,][]{plummer2006coda} should not be applied to the discrete model-indexing variable $z$ because they depend on the arbitrary numerical labels used for the models. 
\cc If such methods are applied nevertheless, the resulting estimate for the effective sample size cannot be interpreted because it is not invariant under permutations of the arbitrary model indices used for the discrete parameter $z$. \sc  
To illustrate this, Figure~\ref{f.neff} shows the distribution of the estimated effective sample size when applying the spectral decomposition available in the R package \texttt{coda} \citep{plummer2006coda} to all 120 permutations of the model indices $(1,\dots,5)$  for a fixed sequence $z^{(t)}$. 
\cc Since this method incorrectly assumes that the discrete variable $z$ is continuous, the estimated effective sample size was not invariant, \sc but varied considerably depending on the specific labeling of the models (gray histogram). 
In contrast, the proposed Markov approach resulted in a well-defined, invariant estimate $\widehat{T}_\text{eff} = 1,921$ (vertical black line) by explicitly accounting for the discreteness of $z$. 

\begin{figure}
\includegraphics[width=10cm]{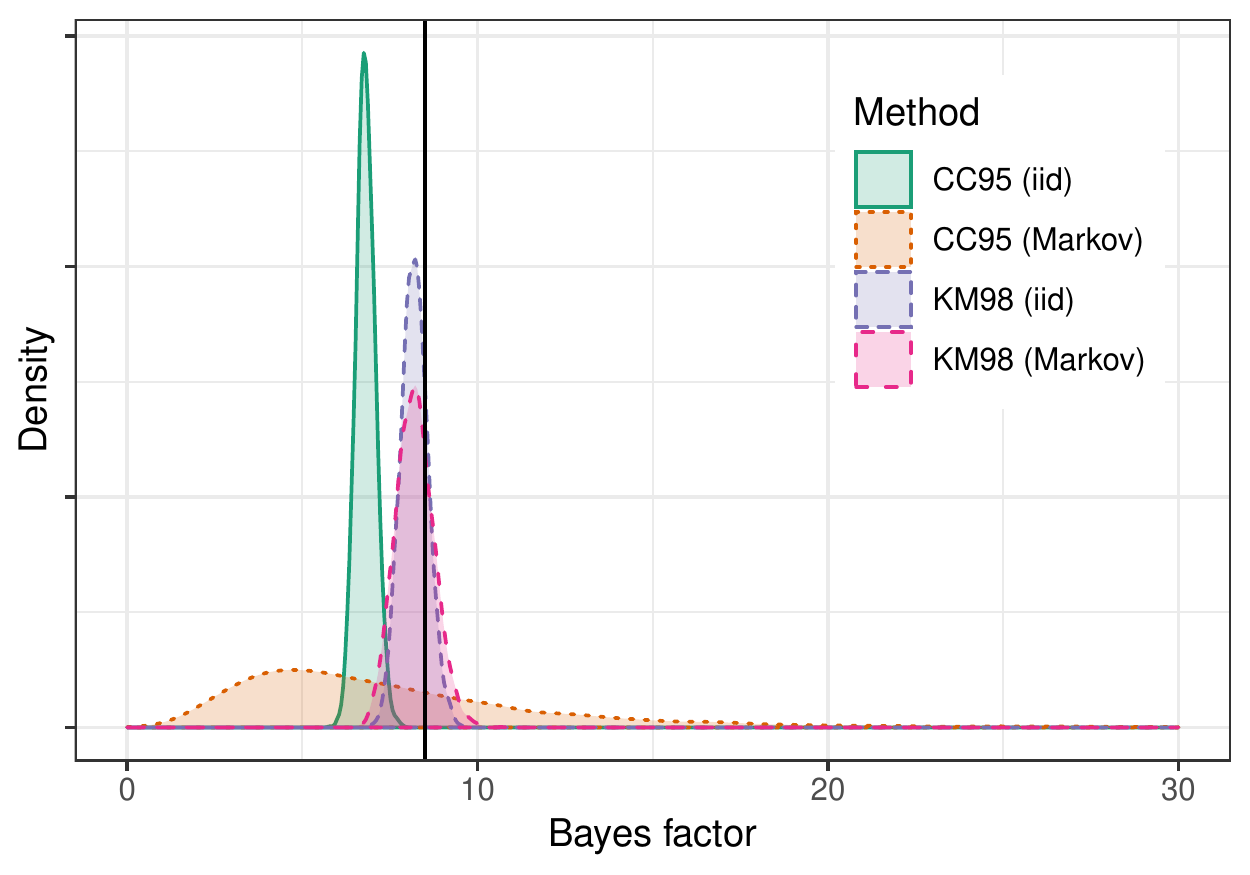}
\caption{
Posterior distribution for the Bayes factor in favor of \cc Model A+B (only main effects) vs. AB (two main effects and interaction). 
The vertical black line shows the target value estimated using two different transdimensional MCMC samplers \sc (CC95 = \citealp{carlin1995bayesian}; KM98 = \citealp{kuo1998variable}).
\cc In contrast to the Markov model, the i.i.d. model incorrectly assumes independence and thus overestimated estimation uncertainty. \sc
}
\label{f.posteriorBF}
\end{figure}

Finally, we show that the posterior samples $\bm\pi^{(t)}$ of the model $\mathcal M^\text{Markov}$ can directly be used to assess the uncertainty of Bayes factor estimates. 
For instance, substantive applications could be interested in testing whether to include the interaction term of condition (A) and treatment (B) in a logistic regression model. 
Given the output of a single MCMC run with 10,000 samples, Figure~\ref{f.posteriorBF} shows the resulting posterior distribution of the Bayes factor $B_\text{A+B,AB}$ in favor for the absence of an interaction. 
Similar to the posterior model probabilities, the i.i.d. approach resulted in overconfidence regarding the estimate and most of the probability mass excluded the correct value 8.51 (approximated with a precision of $\text{SD} = 0.020$). 
In contrast, the Markov approach corrected for dependencies in the samples $z^{(t)}$ and included the correct value. 
The same pattern emerged across the 500 replications, that is, the mean estimated SD of the Bayes factor approximated the corresponding empirical SD of the Bayes factor estimates (KM98: 0.56 vs. 0.60; CC95: 74.7 vs. 114.3). 
When using transdimensional MCMC, Bayes factors cannot be expected to be reliably estimated if models are never or very infrequently sampled (e.g., Model 1 in Table~\ref{t.logistic}). 
For instance, the Bayes factor $B_{\text{A},\text{B}} \approx 43.8$ was estimated very imprecisely even in the KM98 approach (mean SD = 13.0; empirical SD = 24.3). 
To obtain more precise Bayes factor estimates in the presence of infrequently sampled models, it is recommended to rerun the transdimensional MCMC chain including only the two relevant models of interest \citep{lodewyckx2011tutorial}. 

\section{Log-Linear Models for a $2^6$ Contingency Table}
\label{s.heart}

The application of the proposed method is also feasible in realistic scenarios with hundreds of sampled models. 
To illustrate this, we reanalyzed the $2^6$ complete contingency table by \citet{edwards1985fast}, which includes six risk factors for coronary heart disease (i.e., smoking, strenuous mental work, strenuous physical work, systolic blood pressure, ratio of $\alpha$ and $\beta$ lipoproteins, and family anamnesis of coronary heart disease). 
We are interested in finding the most parsimonious log-linear model, which accounts for the cell frequencies $y_j$ of cell $j$ ($j = 1,\dots,2^6$) by assuming a Poisson distribution with mean $\mu_j$ and 
\begin{equation}
\log \mu_j = \phi + \bm x_j^\top \bm \beta,
\end{equation}
where $\phi$ is the intercept, $\bm\beta$ the vector of regression parameters, and $\bm x_j^\top$ the (transposed) design vector, which selects the elements of $\bm\beta$ included for modeling cell $j$. 
Here, we consider the class of hierarchical log-linear models that only allow the inclusion of an interaction if the corresponding marginal effects and lower interaction terms are included in the model as well \citep[e.g.,][]{overstall2014default}.

To select between all 7.8 million possible hierarchical log-linear models \citep{dellaportas1999markov}, we used the reversible jump algorithm proposed by \citet{forster2012reversible}, which is implemented in the R package \texttt{conting} \citep{overstall2014conting}. 
Assuming a unit information prior \citep{ntzoufras2003bayesian}, we sampled 100,000 iterations, discarded 10,000 as burn-in, and applied the proposed Markov chain method by drawing $\cc R=5,000$ samples for the posterior model probabilities \cc of the $I^*$ sampled models. \sc
To assess whether the estimated uncertainty accurately quantifies sampling variability, we ran 200 replications initialized with randomly chosen models. 

Across replications, 5,805 models were sampled (on average, 562.7 per replication). 
Table~\ref{t.conting} shows the results for the 10 models with the highest posterior probabilities.
\cc All of these 10 models included the six main effects (A: smoking, B: strenuous mental work, C: strenuous physical work, D: systolic blood pressure, E: ratio of $\alpha$ and $\beta$ lipoproteins, F: family anamnesis of coronary heart disease) and the first-order interactions AC, AD, AE, BC, and DE, but differed with respect to including the remaining interactions.  \sc 
\cc Despite the large number of iterations, the estimation uncertainty (i.e., the posterior SD)  of the posterior model probabilities was relatively large, \sc indicating that the samples $z^{(t)}$ were autocorrelated to a substantial degree. 
This is also reflected by the effective sample size, which was estimated to be $\widehat T_\text{eff} = 4,259$ on average ($\text{SD} = 181$), approximately $5\%$ of the number of iterations after burn-in.

Table~\ref{t.conting} also shows means and standard deviations of the \cc sampled model rank $\tau$ \sc for the models with the highest posterior probability, indicating that estimation uncertainty (i.e., the posterior SD) increased for models with smaller posterior probabilities. 
Moreover, the proportion of replications is shown for which the sampled rank $\tau$ was identical to the model index ($\tau=\#$) and smaller than or equal to 10 ($\tau \leq 10$). 
According to these proportions, exact ranks were estimated precisely only for the two best models, whereas the set of the 10 models with highest posterior probabilities was relatively stable across posterior samples (with the exception of model 10). 
Importantly, the Markov approach provided mean estimated probabilities $\overline{P(\tau = \#)}$ and $\overline{P(\tau \leq 10)}$ that matched the corresponding empirical proportions across replications.

Note that these results regarding estimation uncertainty are in line with our expectations --- if models have small posterior probabilities, they are also sampled infrequently, which in turn results in estimation uncertainty. 
To quantify this variability, the proposed Markov chain approach provides an  estimate for the achieved precision to assess the quality of the results and to find an appropriate stopping rule for MCMC sampling.

\begin{table}[ht]
\caption{
Models with the highest posterior probability for the $2^6$ contingency table.
}
\label{t.conting}
\hspace{-1.3cm}
\scalebox{0.85}{
\begin{threeparttable}
\begin{tabular}{rlrrrrrrrrrrr}
\hline
& & \multicolumn{4}{c}{Posterior model probabilities ${\bm \pi}$} &
    \multicolumn{7}{c}{\cc Rank $\tau$} \\
\cmidrule(lr){3-6}\cmidrule(lr){7-13}
\# & Model & 
	$\text{Mean}(\bm{\hat\pi})$ &  
	$\text{SD}(\bm{\hat\pi})$ &
	\cc $\overline{\text{SD}}_\text{iid}$ &  
	\cc $\overline{\text{SD}}_\text{Markov}$  & 
	\cc $\text{Mean}(\tau)$ & 
	\cc $\text{SD}(\tau)$ & 
	\cc $\overline{\text{SD}(\tau)}$ & 
	\cc $\tau = \#$ &
	\cc $\overline{P(\tau = \#)}$ & 
	\cc $\tau \leq 10$ &
	\cc $\overline{P(\tau\leq 10)}$  \\ 
 \hline
  1 & CE & 18.78 & 1.34 & 0.13 & 1.02 & 1.00 & 0.00 & 0.03 & 1.00 & 1.00 & 1.00 & 1.00 \\ 
  2 & BE & 11.92 & 0.94 & 0.11 & 0.84 & 2.00 & 0.00 & 0.04 & 1.00 & 1.00 & 1.00 & 1.00 \\ 
  3 & BE + CE & 7.12 & 1.11 & 0.09 & 0.43 & 3.34 & 0.61 & 0.37 & .72 & .78 & 1.00 & 1.00 \\ 
  4 & BF + CE & 6.57 & 1.20 & 0.08 & 0.52 & 3.94 & 0.84 & 0.42 & .71 & .75 & 1.00 & 1.00 \\ 
  5 & BE + BF & 4.20 & 0.85 & 0.07 & 0.41 & 5.42 & 1.59 & 0.21 & .92 & .93 & .96 & .99 \\ 
  6 & CE + EF & 2.77 & 0.50 & 0.06 & 0.33 & 6.80 & 1.71 & 0.58 & .62 & .65 & .94 & 1.00 \\ 
  7 & BE + BF + CE & 2.53 & 0.60 & 0.05 & 0.24 & 8.24 & 5.64 & 0.54 & .58 & .66 & .92 & 1.00 \\ 
  8 & CE + ADE & 1.88 & 0.30 & 0.05 & 0.25 & 8.72 & 1.35 & 0.80 & .47 & .56 & .95 & .95 \\ 
  9 & BE + EF & 1.76 & 0.38 & 0.04 & 0.26 & 9.43 & 3.21 & 0.88 & .45 & .54 & .92 & .93 \\ 
  10 & BE + ADE & 1.19 & 0.22 & 0.04 & 0.19 & 12.05 & 3.11 & 1.40 & .32 & .39 & .39 & .56 \\ 
   \hline
\end{tabular}
\begin{tablenotes}
\note
\cc All of the 10 models include the six main effects, \sc A: smoking, B: strenuous mental work, C: strenuous physical work, D: systolic blood pressure, E: ratio of $\alpha$ and $\beta$ lipoproteins, F: family anamnesis of coronary heart disease, and the first-order interactions AC, AD, AE, BC, and DE. 
Posterior model probabilities ${\bm\pi}$ are shown in percent. 
$\text{Mean}(\bm{\hat\pi})$, $\text{SD}(\bm{\hat\pi})$, $\text{Mean}(\tau)$, and $\text{SD}(\tau)$ were computed across 200 replications. 
\cc The columns $\tau = \#$ and $\tau \leq 10$ refer to the proportion of replications for which the model rank $\tau$ was (a) equal to the model index $\#$ or (b) smaller than or equal to 10. \sc
\end{tablenotes}
\end{threeparttable}
}
\end{table}

\section{Conclusion}

We proposed a novel approach for estimating the precision of transdimensional MCMC output. 
Essentially, a first order Markov model is fitted to the observed model-indexing variable $z^{(t)}$ to quantify estimation uncertainty of the corresponding stationary distribution. 
We showed that the method accounts for autocorrelation in a given sequence $z^{(t)}$ and provides a good assessment of estimation uncertainty. 
Importantly, the method does not require output of multiple independent MCMC chains and thus reduces the computational costs for adaption and burn-in. 
Besides being useful for transdimensional MCMC output, the method provides an estimate of the precision and effective sample size of discrete parameters in MCMC samplers in general. 
Thereby, researchers can easily decide whether the obtained precision is sufficiently high for substantive applications of interest.

\section*{Acknowledgments}

Daniel W. Heck was supported by the research training group \textit{Statistical Modeling in Psychology} (GRK~2277), funded by the German Research Foundation (DFG).

\newpage

\cc
\section{Appendix: Estimating the Shape Parameters of a Dirichlet Distribution}

In the following, we outline the fixed-point algorithm proposed by \citet{minka2000estimating} to estimate the vector of shape parameters $\bm\alpha = (\alpha_1,\dots, \alpha_I)^\top$ of a Dirichlet distribution.
Given a set of $R$ probability vectors $\bm\pi^{(r)}$ (in the proposed method, these are the derived samples of the posterior model probabilities), the likelihood function of the shape parameters $\bm\alpha$ is
\begin{equation}
L\left(\bm\alpha\right) = 
\prod_{r=1}^R \left[
	\frac{\Gamma\left(\sum_i \alpha_i\right)}
	 	{\prod_i \Gamma(\alpha_i)} 
	\prod_i \left(\pi_i^{(r)}\right)^{\alpha_i - 1} 
\right].
\end{equation}
To maximize this likelihood function, \citet{minka2000estimating} developed an efficient fixed-point algorithm and proved its convergence to the unique maximum likelihood estimate $\hat{\bm\alpha}$.
The computational steps are outlined in Algorithm~\ref{a.dirichlet}.
At its core, the current estimates $\alpha_i$ are updated in line~8 by using the digamma function $\Psi$ and its inverse $\Psi^{-1}$.
As remarked by \citet{minka2000estimating}, the algorithm converges very fast even for a large number of shape parameters $I$ (e.g., 80 milliseconds on an Intel\textsuperscript{\textregistered} i7-7700HQ for $I=1,000$).


\begin{algorithm}[] \setstretch{1.5}
\caption{Estimating the shape parameters $\bm\alpha$ of a Dirichlet distribution.\cc}
\label{a.dirichlet} 
\begin{algorithmic}[1]  
\Procedure{Dirichlet Estimation (Minka, 2000)}{}
\State Compute $\bm\mu$: $\mu_i \gets \frac 1 R \sum_{r=1}^R \log \pi_i^{(r)}$
\State Set starting values $\bm\alpha$ with $\alpha_i > 0$ for all $i=1,\dots,I$
\State Set absolute tolerance $\epsilon > 0$ and  $\delta \gets \infty$
\While { $\delta > \epsilon$ }
	\State $\bm\alpha' \gets \bm\alpha$
	\For {$i = 1,\dots,I$}
		\State $\alpha_i \gets \large\Psi^{-1}\left( \large\Psi(\sum_j \alpha_j) + \mu_i \right)$
	\EndFor
	\State $\delta \gets ||\bm\alpha' - \bm\alpha||$
\EndWhile
\State \Return $\bm\alpha$ 
\EndProcedure
\end{algorithmic}
\end{algorithm}

\end{document}